\documentclass[aps,prl,twocolumn,superscriptaddress,footinbib]{revtex4}
\pdfoutput=1
\usepackage{amsmath}
\usepackage{graphicx}
\usepackage{amssymb}
\usepackage{enumitem}
\usepackage{mathrsfs}
\usepackage[linktocpage=true]{hyperref}

\usepackage{color}

\def\nn{\nonumber}
\newcommand{\ii}{\mathrm{i}}
\newcommand{\apar}{a}
\newcommand{\Z}{\mathbb{Z}}

\begin{document}

\title{D3-branes wrapped on a spindle}

\author{Pietro Ferrero}
\affiliation{Mathematical Institute, University of Oxford, Woodstock Road, Oxford, OX2 6GG, U.K.}
\author{Jerome P. Gauntlett}
\affiliation{Blackett Laboratory, Imperial College, Prince Consort Road, London, SW7 2AZ, U.K.}
\author{Juan Manuel P\'erez Ipi\~na}
\affiliation{Mathematical Institute, University of Oxford, Woodstock Road, Oxford, OX2 6GG, U.K.}
\author{Dario Martelli}
 \affiliation{Dipartimento di Matematica, 
 Universit\`a di Torino, 
Via Carlo Alberto 10, 10123 Torino, Italy}
\affiliation{INFN, Sezione di Torino \&   Arnold-Regge Center,
 Via Pietro Giuria 1, 10125 Torino, Italy}
\author{James Sparks}
\affiliation{Mathematical Institute, University of Oxford, Woodstock Road, Oxford, OX2 6GG, U.K.}

\begin{abstract}
\noindent  We construct supersymmetric $AdS_3\times \Sigma$ solutions 
of minimal gauged supergravity in $D=5$, where $\Sigma$ is a two-dimensional orbifold known as a spindle. 
Remarkably, these uplift on $S^5$, or more generally on any regular Sasaki-Einstein manifold, to smooth solutions of 
type IIB supergravity. The solutions are dual to $d=2$, $\mathcal{N}=(0,2)$ SCFTs and we show that the
central charge for the gravity solution agrees with a field theory calculation associated with D3-branes wrapped on $\Sigma$.

\end{abstract}

\maketitle

\section{Introduction}\label{sec:intro}

Important insights into strongly coupled supersymmetric conformal field theories (SCFTs)  
can be obtained by realising them as the renormalization group fixed points of compactifications of higher-dimensional 
field theories. Such SCFTs can be constructed by starting with 
the field theories arising on the worldvolumes of branes in string theory or M-theory, wrapping them 
on a compact manifold $\Sigma$, and then flowing to the infrared (IR). 

In favourable circumstances these configurations can be described within the AdS/CFT correspondence
by brane solutions of $D=10/11$ supergravity. Such solutions have a boundary of the form $AdS_{d+1}\times M$, where $M$ is a compact manifold, which describes the ultraviolet (UV) of the SCFT in $d$ spacetime dimensions. They also have near horizon geometries of the schematic 
form $AdS_{d+1-n}\times\Sigma\times M$, which describes the SCFT in $d-n$ dimensions arising in the IR, where $\Sigma$ has dimension $n$. After a Kaluza-Klein reduction of the $D=10/11$ supergravity theory on $M$, to obtain a gravity theory in $d+1$ spacetime dimensions,
these solutions may be viewed as black branes in $AdS_{d+1}$ with horizon $\Sigma$.

Such solutions were first constructed in the foundational work \cite{Maldacena:2000mw} and describe M5-branes and D3-branes wrapping 
Riemann surfaces of  constant curvature, with genus $g>1$. Subsequently there have been many generalisations 
including wrapping manifolds of higher dimension, relaxing the constant curvature condition, allowing for punctures etc.
(e.g. \cite{Gauntlett:2000ng,Gaiotto:2009gz,Anderson:2011cz}).
In all these developments, supersymmetry is preserved by demanding that the field theory arising on the brane worldvolume is  
``topologically twisted" \cite{Witten:1988ze,Bershadsky:1995qy}. This involves a specific coupling both to the metric on $\Sigma$
and to external R-symmetry gauge fields. An important consequence of this twisting is that the Killing spinors
preserved by the solution are independent of the coordinates of $\Sigma$.

Here we discuss a class of supersymmetric solutions which have fundamentally new features. We present $AdS_3\times \Sigma$ solutions of $D=5$ minimal gauged supergravity which we
interpret as the near horizon limit of black brane solutions associated with D3-branes wrapped on $\Sigma$.  The first new feature is that
supersymmetry is \emph{not} realised by a topological twist. 
The second is that
$\Sigma$ is not a compact manifold but an \emph{orbifold} \footnote{An orbifold is locally modelled on open subsets of $\mathbb{R}^n/\Gamma$, where $\Gamma$ are finite groups.}. More specifically, we will consider the weighted projective space $\Sigma=\mathbb{WCP}^1_{[n_-,n_+]}$, also known as a spindle. 
This is topologically a two-sphere but with conical deficit angles $2\pi (1-{1}/{n_\mp})$ at the poles, specified by
two coprime positive integers $n_\mp$, with $n_-\ne n_+$
\footnote{$\mathbb{WCP}^1_{[n_-,n_+]}$ is a bad orbifold in the sense that it is not possible to move to a covering space that is a manifold. 
It does not admit a metric of constant curvature.}.

Remarkably, after uplifting the $AdS_3\times \Sigma$ solutions on specific
Sasaki-Einstein five-manifolds, $SE_5$, to obtain solutions of type IIB supergravity, they become completely 
smooth \footnote{The local $D=5$ solutions were independently presented in \cite{Kunduri:2006uh}, who also made a 
local connection with the solutions of~\cite{Gauntlett:2006af}.}.
Moreover, the resulting $AdS_3\times \mathcal{M}_7$ solutions are precisely those 
 of~\cite{Gauntlett:2006af}. Our construction suggests that the $d=2$, $\mathcal{N}=(0,2)$ 
SCFTs dual to the $AdS_3\times \mathcal{M}_7$ solutions of \cite{Gauntlett:2006af}
arise from compactifying $d=4$, $\mathcal{N}=1$ SCFTs on a spindle, where the $d=4$ theories are dual to 
$AdS_5\times SE_5$. This includes the case of $\mathcal{N}=4$ SYM, 
where $SE_5=S^5$. We make a precision test of this interpretation:
we compute the central charge and superconformal R-symmetry 
of the $d=2$ field theories using anomaly polynomials and $c$-extremization \cite{Benini:2012cz} and find exact agreement with 
the gravity result \cite{Gauntlett:2006af}.

\section{$D=5$ solutions}\label{sec:soln}

The equations of motion for $D=5$ minimal gauged supergravity \cite{Gunaydin:1983bi} are given by
\begin{align}\label{EOM}
\begin{split}
R_{\mu\nu} & = -4g_{\mu\nu} + \tfrac{2}{3}F_{\mu\rho}F_{\nu}^{\ \rho} - \tfrac{1}{9}g_{\mu\nu} F_{\rho\sigma}F^{\rho\sigma}\, ,\\
d *F & =  -\tfrac{2}{3}F\wedge F\, ,
\end{split}
\end{align}
where $F=dA$, with $A$ the Abelian R-symmetry gauge field. 
A solution is supersymmetric if it 
admits a Killing spinor satisfying
\begin{align}\label{KSE}
\left[\nabla_\mu - \tfrac{\ii}{12}(\Gamma_\mu^{\;\; \nu\rho}-4\delta_\mu^\nu \Gamma^\rho)F_{\nu \rho} -\tfrac{1}{2}\Gamma_\mu - \ii 
A_\mu\right]\epsilon = 0 \, ,
\end{align}
where $\epsilon$ is a Dirac spinor and $\{\Gamma_\mu,\Gamma_\nu\}=2g_{\mu\nu}$. 

The supersymmetric solution of interest is given by
\begin{align}\label{soln}
ds^2_5  = \frac{4y}{9}ds^2_{AdS_3} + ds^2_\Sigma\, ,\qquad
A  = \frac{1}{4}\left(1-\frac{\apar}{y}\right)dz\,.
\end{align}
Here $ds^2_{AdS_3}$ is a unit radius metric on $AdS_3$, 
while
\begin{align}\label{Sigmametric}
ds^2_\Sigma =  \frac{y}{q(y)}dy^2 + \frac{q(y)}{36y^2}dz^2\, ,
\end{align}
is the metric on the horizon, $\Sigma$, and
\begin{align}
q(y) = 4y^3 - 9y^2 + 6\apar y - \apar^2\, ,
\end{align}
with $\apar$ a constant. 

Assuming $\apar\in (0,1)$ the three 
roots $y_i$ of $q(y)$ are all real and positive. 
Defining $y_1<y_2<y_3$, we then take $y\in[y_1,y_2]$ 
to obtain a positive definite metric \eqref{Sigmametric} on $\Sigma$.
However, as $y$ approaches $y_1$ and $y_2$ it is not possible to remove the conical 
deficit singularities at both roots by a single choice 
of period $\Delta z$ for $z$, to obtain a smooth two-sphere. Instead we find that if
\begin{align}
\begin{split}
\apar & = \frac{(n_--n_+)^2(2n_-+n_+)^2(n_-+2n_+)^2}{4(n_-^2+n_- n_+ + n_+^2)^3}\, , \\
\Delta z & = \frac{2(n_-^2+n_- n_+ + n_+^2)}{3n_-n_+(n_-+n_+)}2\pi \, ,
\end{split}
\end{align}
then $ds^2_\Sigma$
is a smooth metric on the orbifold $\Sigma=\mathbb{WCP}^1_{[n_-,n_+]}$. 
Specifically, there  are conical deficit angles $2\pi (1-{1}/{n_\mp})$ 
at $y=y_1$, $y_2$, respectively, where $n_\pm$ are arbitrary coprime positive integers with 
$n_->n_+$. 

Note that there is magnetic 
flux through $\Sigma$:
\begin{align}\label{Fflux}
\frac{1}{2\pi} \int_\Sigma F = \frac{n_--n_+}{2n_-n_+}\, .
\end{align}
This may be contrasted with the Euler number:
\begin{align}\label{Euler}
\chi(\Sigma) = \frac{1}{4\pi}\int_{\Sigma} R_\Sigma \mathrm{vol}_\Sigma = \frac{n_-+n_+}{n_-n_+}\, ,
\end{align}
where $R_\Sigma$ is the Ricci scalar of $\Sigma$, and $\mathrm{vol}_\Sigma$ is its volume form
\footnote{In general the integral of the curvature for a
$U(1)$ gauge field on $\mathbb{WCP}^1_{[n_-,n_+]}$ is necessarily $2\pi/(n_-n_+)$ times an integer e.g. \cite{Ferrero:2020twa}. 
This makes the gauge field $A$, with flux given by \eqref{Fflux}, a spin$^c$ gauge field: 
$n_--n_+$ is even/odd precisely when $n_-+n_+$ is even/odd, so that spinor fields with unit charge under $A$ 
are always globally well-defined.}.

To solve \eqref{KSE} we write $\Gamma^a=\gamma^a\otimes\sigma^3$, for $a=0,1,2$ with $\gamma^0=-\ii\sigma^2$, $\gamma^1=\sigma^1$, $\gamma^2=\sigma^3$
and $\Gamma^3=1\otimes\sigma^2$, $\Gamma^4=1\otimes \sigma^1$, where $\sigma^i$ are Pauli matrices. We then write $\epsilon=\vartheta\otimes\chi$
with $\vartheta$ a Killing spinor for $AdS_3$ satisfying $\nabla_{a}\vartheta=\frac{1}{2}\,\gamma_a\,\vartheta$.
The two-component spinor $\chi$ on the spindle is given by
\begin{align}\label{2dspinor}
\chi=
\left(\frac{\sqrt{q_1(y)}}{\sqrt{y}},\ii\frac{ \sqrt{q_2(y)}}{\sqrt{y}}\right)\, ,
\end{align}
where
\begin{align}
q_1(y) = -a+2 y^{3/2}+3 y\,,\,  \, q_2(y) =a+2 y^{3/2}-3 y\, ,
\end{align}
which satisfy $q(y)=q_1(y)q_2(y)$. In contrast to the topological twist, this spinor depends on the coordinates of $\Sigma$. 
Moreover, as shown in \cite{Ferrero:2020twa}, the spinor is in fact a section of a non-trivial bundle over $\Sigma$.
Note that the gauge choice 
used in \eqref{soln} has been fixed by requiring $\epsilon$ to be independent of $z$.

\section{Uplift to IIB string theory}\label{sec:uplift}

Any supersymmetric solution to \eqref{EOM} uplifts (locally) to type IIB supergravity via 
\cite{Buchel:2006gb}:
\begin{align}\label{uplift}
\begin{split}
ds^2_{10} & = L^2\left[ds^2_5 + (\tfrac{1}{3}d\psi +\sigma + \tfrac{2}{3}A)^2 + ds^2_{KE_4}\right]\\
g_sF_5 & =  L^4\Big[4\mathrm{vol}_5 - \tfrac{2}{3}*_5 F\wedge J  \\
& + (2J\wedge J -\tfrac{2}{3}F\wedge J)\wedge (\tfrac{1}{3}d\psi +\sigma + \tfrac{2}{3}A)\Big]\, .
\end{split}
\end{align}
Here $F_5$ is the self-dual five-form,
$g_s$ is the string coupling constant, 
and $L>0$ is a length scale that is fixed by flux quantization.
$KE_4$ is an arbitrary positively curved  K\"ahler-Einstein four-manifold 
with K\"ahler form $J$, normalized so that the Ricci form is  $\mathcal{R}=6J$, 
and $\sigma$ is a local one-form with $d\sigma = 2J$.

Substituting \eqref{soln} into 
\eqref{uplift} we find that the $D=10$ metric may be written as
\begin{align}
ds^2_{10} = \frac{4}{9}L^2y\left[ds^2_{AdS_3} + ds^2_{\mathcal{M}_7}\right]\ ,
\end{align}
where $\mathcal{M}_7$ is a compact seven-manifold. This
is the same solution of type IIB
supergravity given in \cite{Gauntlett:2006af}.

It was shown in \cite{Gauntlett:2006af} that $\mathcal{M}_7$ is the total space of a 
Lens space $S^3/\mathbb{Z}_q$ fibration over the $KE_4$, where 
the twisting is parametrized by another positive integer $p$. The Lens space 
fibre has coordinates $y,z,\psi$.
In terms of our parameters $n_\pm$ we identify
\begin{align}\label{pandq}
p = kn_+\, , \quad q = \frac{k}{I}(n_--n_+)\, ,
\end{align}
where $p,q\in\mathbb{N}$ are coprime.
The Fano index, $I$, is the largest positive integer for which 
$\int_S c_1/I\in \mathbb{Z}$, for all two-cycles $S$ in the $KE_4$,
where $c_1=[\mathcal{R}/2\pi]\in H^2(KE_4,\mathbb{Z})$.
We have also defined
\begin{align}\label{kone}
k = \mathrm{hcf}(I,p)\, ,
\end{align}
and identify $\psi$ with period $\Delta \psi$ given by 
\footnote{The torus parametrised by $(z,\psi)$ is the same as \cite{Gauntlett:2006af}. We have
made identifications on $z$ and $\psi$ in the opposite order to \cite{Gauntlett:2006af}, so
\eqref{delpsi} is not the same but $\Delta\psi\Delta z$ is.}
\begin{align}\label{delpsi}
\Delta \psi = \frac{2\pi I}{k}\, .
\end{align}
At a fixed point in the $D=5$ spacetime, the 
internal five-dimensional metric 
$ds^2_{SE_5}=(\tfrac{1}{3}d\psi+\sigma)^2 + ds^2_{KE_4}$ in
\eqref{uplift} is then a {\it regular} Sasaki-Einstein 
manifold, which is simply-connected when $k=1$. 

To obtain a string theory background one must 
also quantize the five-form flux $F_5$ through all five-cycles in $\mathcal{M}_7$. This was carried out in 
\cite{Gauntlett:2006af}. We define the integers
\begin{align}
M = \int_{KE_4} c_1\wedge c_1 = \frac{1}{4\pi^2}\int_{KE_4}\mathcal{R}\wedge \mathcal{R}\, ,
\end{align}
and $h=\mathrm{hcf}(M/I^2,q)$. Then
if we choose $L$ to satisfy 
\begin{align}
\frac{L^4}{g_s\ell_s^4} = \frac{108\pi}{I^3h}k^2n_+n_- n\, ,
\end{align}
where $\ell_s$ is the string length and $n\in\mathbb{N}$, then  one finds that
$1/(2\pi \ell_s)^4\int_D F_5\in\Z$, for all 
five-cycles $D$ in $\mathcal{M}_7$. 

There is a finite set of choices for the positively curved $KE_4$.
If $KE_4=\mathbb{CP}^2$ then $I=3$, $M=9$. For this case,
$k=1$ gives $SE_5=S^5$ as the internal space  while $k=3$ gives $S^5/\mathbb{Z}_3$. 
If $KE_4=S^2\times S^2$ we have $I=2,M=8$. Now $k=1$ gives $SE_5=T^{1,1}$, while $k=2$ gives $SE_5=T^{1,1}/\mathbb{Z}_2$.
Finally, for $KE_4=dP_m$, $3\leq m\leq 8$, where $dP_m$ is a del Pezzo surface, we have $I = 1$, $M = 9 -m$. 

A key observation is that $\mathcal{M}_7$ in the $AdS_3$ solutions of 
\cite{Gauntlett:2006af} may also be viewed as $SE_5$ fibrations 
over $\Sigma=\mathbb{WCP}^1_{[n_-,n_+]}$. We can begin 
with any weighted projective space, with weights $n_->n_+$, and 
then define $p,q\in\mathbb{Z}$ via \eqref{pandq}, where we also define
\begin{align}\label{ktwo}
k = \frac{I}{\mathrm{hcf}(I,n_--n_+)}\, .
\end{align}
With this definition, $p$ and $q$ are manifestly coprime,  and 
one can check that \eqref{kone} is equivalent to \eqref{ktwo}. 
With this perspective, we can calculate the flux of $F_5$ through the $SE_5$ fibre:
\begin{align}\label{Nflux}
N \equiv  \frac{1}{(2\pi \ell_s)^4}\int_{SE_5} F_5 = \frac{M}{I^2h}kn_+n_- n \in \mathbb{N}\, .
\end{align}
Notice that for a given spindle, specified by $n_\pm$, and a given choice of $KE_4$ we only
get a smooth type IIB solution for $k$ as in \eqref{ktwo} and hence a specific $SE_5$. E.g. if
$KE_4=\mathbb{CP}^2$ and $n_+=2$, then for $n_-=3,7,\ldots$ and $n_-=5,9,\ldots $ 
we can uplift on $S^5/\mathbb{Z}_3$ and $S^5$, respectively. 

The central charge is given by $c = {3L}/{2G_{(3)}}$, where $G_{(3)}$ 
is the Newton constant  obtained by compactifying type IIB supergravity on 
$\mathcal{M}_7$ \cite{Brown:1986nw}.
We can rewrite the result of \cite{Gauntlett:2006af} as
\begin{align}\label{cgravity}
c = \frac{4(n_--n_+)^3}{3n_-n_+(n_-^2+n_-n_++n_+^2)}a_{4d}\, ,
\end{align}
where 
\begin{align}\label{a4d}
a_{4d} \equiv  \frac{\pi^2 N^2}{4\mathrm{vol}(SE_5)}\, .
\end{align}

In \cite{Gauntlett:2006af} the dual $d=2$, $\mathcal{N}=(0,2)$ SCFTs
were not identified, but our $D=5$ construction of the solutions,
together with the flux condition \eqref{Nflux}, leads to a conjecture. Begin with the $d=4$ SCFT
dual to $AdS_5\times SE_5$, describing $N$ D3-branes at the Calabi-Yau 
three-fold singularity with conical metric $dr^2 + r^2ds^2_{SE_5}$. 
The large $N$ $a$-central charge of this theory is precisely given by 
$a_{4d}$ in \eqref{a4d} \cite{Gubser:1998vd}.  
One then compactifies that theory on $\Sigma=\mathbb{WCP}^1_{[n_-,n_+]}$, 
with a background R-symmetry gauge field with magnetic flux \eqref{Fflux}. 
The solutions we have described suggest the theory flows 
to a $d=2$, $\mathcal{N}=(0,2)$ SCFT in the IR, and we will 
give evidence for this below by computing the 
central charge via a field theory calculation. 

The $U(1)_R$ symmetry of the $(0,2)$ theory dual to the $AdS_3\times\mathcal{M}_7$ solutions
is realized by a Killing vector, $R_{2d}$, on $\mathcal{M}_7$.
Using the results of \cite{Kim:2005ez,Gauntlett:2006ns} we deduce
\begin{align}\label{R2d}
R_{2d}=2\partial_\psi + \frac{3n_-n_+(n_-+n_+)}{n_-^2+n_-n_++n_+^2}\partial_\varphi\, .
\end{align}
Here we have defined $\varphi = \frac{2\pi z}{\Delta z}$
so that $\Delta\varphi=2\pi$, and $\partial_\varphi$ generates 
the $U(1)$ isometry of the weighted projective space $\Sigma$, 
which we shall refer to as $U(1)_J$. 
Note that the Killing 
spinor on the $SE_5$ has unit charge under $R_{4d}=2\partial_\psi$, 
which may therefore be identified with the superconformal $U(1)_R$ symmetry 
of the $d=4$ SCFT before compactification on $\Sigma$. 
In other words, equation \eqref{R2d} states that the $d=4$ R-symmetry mixes with $U(1)_J$
in flowing to the $d=2$ R-symmetry in the IR. 
We shall also recover \eqref{R2d} from a field theory calculation in the next section. 

\section{$d=4$ SCFTs on $\Sigma$}

We begin with a general $d=4$ SCFT with anomaly polynomial given by the 6-form
\begin{align}
\mathcal{A}_{4d}=\frac{\mathrm{tr}\, R^3}{6} c_1(R_{4d})^3 - \frac{\mathrm{tr}\, R}{24} c_1(R_{4d})p_1(TZ_6)\, .
\end{align}
As is standard, in (even) dimension $d$ the anomaly polynomial 
is a $(d+2)$-form on an abstract $(d+2)$-dimensional space, here called $Z_6$.
$c_1(R_{4d})$ denotes the first Chern class of the $d=4$ superconformal $U(1)_R$ symmetry bundle over $Z_6$, 
and $p_1$ denotes the first Pontryagin class. 
The trace is over Weyl fermions when the theory has a Lagrangian description, and 
in any case 
we may always write 
\begin{align}
\mathrm{tr}\, R^3 = \frac{16}{9}(5a_{4d}-3c_{4d})\, , \quad \mathrm{tr}\, R = 16 (a_{4d}-c_{4d})\, ,
\end{align}
in terms of the central charges $a_{4d}$, $c_{4d}$. 
We focus on the large $N$ limit in which 
$a_{4d}=c_{4d}$ to leading order, and hence 
\begin{align}\label{A4d}
\mathcal{A}_{4d}  = \frac{16 a_{4d}}{27} c_1(R_{4d})^3 \quad (\mbox{at large $N$})\, .
\end{align}

We now compactify the $d=4$ theory on $\Sigma=\mathbb{WCP}^1_{[n_-,n_+]}$, 
with magnetic flux  \eqref{Fflux} for the $d=4$ R-symmetry gauge field. 
The resulting $d=2$ anomaly polynomial will then capture the 
right-moving central charge $c_r$, 
but crucially we need to include the 
$U(1)_J$ global symmetry in $d=2$ that comes from the isometry of $\Sigma$. 
Geometrically, this involves taking $Z_6$ to be the total space of a 
$\Sigma$ fibration over a four-manifold $Z_4$ \cite{Hosseini:2020vgl}. More precisely, 
we let ${J}$ be a  $U(1)$ bundle over $Z_4$,
with connection corresponding to a background 
gauge field $A_J$ for the $d=2$ $U(1)_J$ global symmetry, 
and then fibre $\Sigma$ over $Z_4$ using the $U(1)_J$ action
and connection $A_J$. 
In practice this amounts to the replacement $d\varphi \mapsto
d\varphi + A_J$.

Incorporating the magnetic flux \eqref{Fflux} into this construction 
amounts to ``gauging'' the $U(1)$ gauge field $A$ in \eqref{soln}, as just 
described. Thus, we define the following connection one-form on $Z_6$:
\begin{align}\label{fibreA}  
\mathscr{A} = \frac{1}{4}\left(1-\frac{\apar}{y}\right) \frac{\Delta z}{2\pi}(d\varphi + A_J) \equiv \rho(y)(d\varphi+A_J)\, ,
\end{align}
where recall that the gauge choice we made is such that the Killing 
spinors are uncharged under the $U(1)_J$ symmetry generated by $\partial_\varphi$. 
This is necessary for the twisting to make sense. 
$\mathscr{A}$ defined by \eqref{fibreA} is a gauge field on $Z_6$, which restricts to the supergravity gauge field $A$ in \eqref{soln}
on each $\Sigma$ fibre. We compute the curvature
\begin{align}
\mathscr{F} = d\mathscr{A} =  \rho'(y) d y \wedge(d\varphi + A_J) + \rho(y)F_J\, ,
\end{align}
where $F_J=dA_J$. 
The one-form $d\varphi+A_J$ is precisely the global angular form 
for the $U(1)$ bundle, and so is globally defined on $Z_6$ away from the 
poles of $\Sigma=\mathbb{WCP}^1_{[n_-,n_+]}$ at $y=y_1,y_2$. 
Moreover, one can verify that $\rho'(y)dy\wedge d\varphi$ vanishes smoothly 
at the poles, where the angular coordinate $\varphi$ is not defined, 
implying that $\mathscr{F}$ is a globally defined closed two-form 
on $Z_6$. By construction the integral of $\mathscr{F}$ over a fibre 
$\Sigma$ of $Z_6$ satisfies \eqref{Fflux}.  More generally, the integrals of wedge products over the fibres are given, for $s\in\mathbb{N}$, by
\begin{align}\label{Fintegrals}
\int_{\Sigma}\left(\frac{\mathscr{F}}{2\pi}\right)^s & = \frac{1}{2^s}\left(\frac{1}{n_+^s}-\frac{1}{n_-^s}\right)\left(-\frac{F_J}{2\pi}\right)^{s-1}\,.
\end{align}

The curvature form $\mathscr{F}$ defines a  $U(1)$ bundle $\mathcal{L}$ over $Z_6$
by taking $c_1(\mathcal{L}) = [\mathscr{F}/2\pi] \in H^2(Z_6,\mathbb{R})$. This is different 
from \cite{Hosseini:2020vgl}, where the $U(1)$ bundle was taken to be the 
tangent bundle to the fibres  
 $T_{\mathrm{fibres}}Z_6$, 
 which gives the Euler class \eqref{Euler}, rather than  
\eqref{Fflux}. 
We note that 
at the poles we have $c_1(\mathcal{L})\mid_{y=y_1,y_2} =-\frac{1}{2n_\pm}c_1(J)$, where we have defined $c_1(J)=[F_J/2\pi] \in H^2(Z_4,\Z)$. 
In the anomaly polynomial we then write
\begin{align}\label{c1}
c_1(R_{4d}) = c_1(R_{2d}) + c_1(\mathcal{L}) \, ,
\end{align}
where $R_{2d}$ is the pull-back of a $U(1)$ bundle over $Z_4$.
Notice that the twisting \eqref{c1} will make sense globally only if the $d=4$ R-charges 
of fields satisfy appropriate quantization conditions, and for gauge-invariant operators 
this is equivalent to the global regularity and flux quantization conditions imposed on the supergravity solutions, 
{\it cf.} the discussion below equation \eqref{Nflux}.

The $d=2$ anomaly polynomial is obtained by integrating $\mathcal{A}_{4d}$ in \eqref{A4d} 
over $\Sigma$. Using  \eqref{Fintegrals} and \eqref{c1} we compute
\begin{align}\label{A2d}
&\mathcal{A}_{2d}  =\frac{2a_{4d}}{27}\Big[
 {12}\Big(\frac{1}{n_+} -\frac{1}{n_-}\Big) c_1(R_{2d})^2\nn\\
&
-6\Big(\frac{1}{n_+^2}-\frac{1}{n_-^2}\Big)c_1(R_{2d}) c_1(J)+\Big(\frac{1}{n_+^3}-\frac{1}{n_-^3}\Big)c_1(J)^2
\Big] \, .
\end{align}
The coefficient of $\tfrac{1}{2}c_1(L_i) c_1(L_j)$ in $\mathcal{A}_{2d}$ is $\mathrm{tr}\, \gamma^3 Q_i Q_j$, 
where the global symmetry $Q_i$ is associated to the $U(1)$ bundle $L_i$ over $Z_4$, and $\gamma^3$ 
is the $d=2$ chirality operator. 
On the other hand, $c$-extremization \cite{Benini:2012cz} 
implies that the $d=2$ superconformal $U(1)_R$ 
extremizes 
\begin{align}\label{ctrial}
c_{\mathrm{trial}} = 3\, \mathrm{tr}\, \gamma^3 R_{\mathrm{trial}}^2\, ,
\end{align}
over the space of possible R-symmetries. We set
\begin{align}\label{Rtrial}
R_{\mathrm{trial}} = R_{2d} + \varepsilon\, J\,,
\end{align}
and extremize the quadratic function of $\varepsilon$ one obtains from 
\eqref{A2d} and \eqref{ctrial}. The extremal  
value is 
\begin{align}\label{epsilon}
\varepsilon_* = \frac{3n_-n_+(n_-+n_+)}{n_-^2+n_-n_++n_+^2}\ .
\end{align}
The right-moving central charge is given by
\eqref{ctrial} evaluated on the superconformal R-symmetry \cite{Benini:2012cz}.
Substituting \eqref{epsilon} into \eqref{ctrial}, \eqref{Rtrial} we find
\begin{align}\label{cr}
c_r = \frac{(n_--n_+)^3}{n_-n_+(n_-^2+n_-n_++n_+^2)}\frac{4a_{4d}}{3}\, .
\end{align}
At leading order in $N$, $c_r=c_l\equiv c$ is the central 
charge of the SCFT, and we see that the field theory 
result \eqref{cr} precisely matches the gravity 
result \eqref{cgravity}. Moreover, the 
R-symmetry \eqref{Rtrial}, with $\varepsilon=\varepsilon_*$, precisely matches the supergravity 
R-symmetry \eqref{R2d}.

\section{Discussion}
Our solutions exhibit a number of new properties, 
raising several directions for future research. 
Firstly, despite having orbifold singularities in $D=5$, when 
uplifted to $D=10$ the solutions are completely regular. 
What type of singularities are permitted in lower-dimensional
supergravity theories that have this property? 
Secondly, it is often claimed that supersymmetry requires a topological twist when branes wrap a compact manifold, so that the ``twisted spinors'' are constant on the manifold. 
Our near horizon solutions are a counterexample and it would be interesting 
to understand this more systematically.
Thirdly, our results suggest there should exist black string solutions which approach $AdS_5$ in the UV and 
$AdS_3\times \Sigma$ in the IR. Such solutions will reveal the precise deformations of the 
$d=4$, $\mathcal{N}=1$ SCFTs that can then flow to the $d=2$, $\mathcal{N}=(0,2)$ SCFTs dual to the $AdS_3\times \mathcal{M}_7$ solutions of \cite{Gauntlett:2006af}, including the way in which the D3-brane wrapping the spindle is preserving supersymmetry. 

In \cite{Ferrero:2020twa} we present analogous supergravity solutions in $D=4$, 
associated with M2-branes wrapped on a ``spinning spindle''. In that case the full 
black hole solution is known and it approaches $AdS_4$ in the UV and $AdS_2\times \Sigma$ in the IR, with a spindle horizon $\Sigma$.  
The $D=4$ black hole is accelerating and this leads to the conical deficit singularities on $\Sigma$. 
Once lifted to $D=11$ these orbifold singularities are removed and
the solutions become completely regular.
Supersymmetry is again not realised by the topological twist for the $AdS_2\times \Sigma$ solution, similar to our $AdS_3\times \Sigma$ solutions. In the UV, the black holes at finite temperature have a conformal boundary consisting of a spindle.
In the supersymmetric and extremal limit, however, the spindle degenerates into ``two halves'', each of which {\it is} associated with a topological twist,
but with a different constant spinor on each component!
It is another fascinating open question to determine how typical such a novel realisation of supersymmetry is for branes wrapping spindles, as well as spaces of higher dimension.

Our results also suggest many questions on the field theory side. When defining a SCFT on a spindle what additional data should be specified at the orbifold points? Considering weakly coupled $\mathcal{N}=4$ SYM theory would be an important first step.
Our holographic analysis shows that SCFTs dual to regular SE manifolds can only be placed on certain spindles and there is an apparent obstruction for
those dual to irregular SE manifolds; why is this? The anomaly polynomial technique we employed is based on smooth manifolds and 
yet it gives a consistent result in the context of the orbifolds we studied, in the large $N$ limit. It would be interesting to justify this more systematically
(also see \cite{Bah:2020jas})
and determine sub-leading contributions.

\section*{Acknowledgments}

\noindent We thank K.~Hristov and J.~Lucietti for comments.
This work was supported in part by STFC grants ST/P000762/1, ST/T000791/1 and 
ST/T000864/1. 
JPG is supported as a KIAS Scholar and as a Visiting Fellow at the Perimeter Institute. 


\begin{thebibliography}{23}%
\makeatletter
\providecommand \@ifxundefined [1]{%
 \@ifx{#1\undefined}
}%
\providecommand \@ifnum [1]{%
 \ifnum #1\expandafter \@firstoftwo
 \else \expandafter \@secondoftwo
 \fi
}%
\providecommand \@ifx [1]{%
 \ifx #1\expandafter \@firstoftwo
 \else \expandafter \@secondoftwo
 \fi
}%
\providecommand \natexlab [1]{#1}%
\providecommand \enquote  [1]{``#1''}%
\providecommand \bibnamefont  [1]{#1}%
\providecommand \bibfnamefont [1]{#1}%
\providecommand \citenamefont [1]{#1}%
\providecommand \href@noop [0]{\@secondoftwo}%
\providecommand \href [0]{\begingroup \@sanitize@url \@href}%
\providecommand \@href[1]{\@@startlink{#1}\@@href}%
\providecommand \@@href[1]{\endgroup#1\@@endlink}%
\providecommand \@sanitize@url [0]{\catcode `\\12\catcode `\$12\catcode
  `\&12\catcode `\#12\catcode `\^12\catcode `\_12\catcode `\%12\relax}%
\providecommand \@@startlink[1]{}%
\providecommand \@@endlink[0]{}%
\providecommand \url  [0]{\begingroup\@sanitize@url \@url }%
\providecommand \@url [1]{\endgroup\@href {#1}{\urlprefix }}%
\providecommand \urlprefix  [0]{URL }%
\providecommand \Eprint [0]{\href }%
\providecommand \doibase [0]{http://dx.doi.org/}%
\providecommand \selectlanguage [0]{\@gobble}%
\providecommand \bibinfo  [0]{\@secondoftwo}%
\providecommand \bibfield  [0]{\@secondoftwo}%
\providecommand \translation [1]{[#1]}%
\providecommand \BibitemOpen [0]{}%
\providecommand \bibitemStop [0]{}%
\providecommand \bibitemNoStop [0]{.\EOS\space}%
\providecommand \EOS [0]{\spacefactor3000\relax}%
\providecommand \BibitemShut  [1]{\csname bibitem#1\endcsname}%
\let\auto@bib@innerbib\@empty
\bibitem [{\citenamefont {Maldacena}\ and\ \citenamefont
  {Nunez}(2001)}]{Maldacena:2000mw}%
  \BibitemOpen
  \bibfield  {author} {\bibinfo {author} {\bibfnamefont {J.~M.}\ \bibnamefont
  {Maldacena}}\ and\ \bibinfo {author} {\bibfnamefont {C.}~\bibnamefont
  {Nunez}},\ }\href {\doibase 10.1142/S0217751X01003937} {\bibfield  {journal}
  {\bibinfo  {journal} {Int.J.Mod.Phys.}\ }\textbf {\bibinfo {volume} {A16}},\
  \bibinfo {pages} {822} (\bibinfo {year} {2001})},\ \Eprint
  {http://arxiv.org/abs/hep-th/0007018} {arXiv:hep-th/0007018 [hep-th]}
  \BibitemShut {NoStop}%
\bibitem [{\citenamefont {Gauntlett}\ \emph {et~al.}(2001)\citenamefont
  {Gauntlett}, \citenamefont {Kim},\ and\ \citenamefont
  {Waldram}}]{Gauntlett:2000ng}%
  \BibitemOpen
  \bibfield  {author} {\bibinfo {author} {\bibfnamefont {J.~P.}\ \bibnamefont
  {Gauntlett}}, \bibinfo {author} {\bibfnamefont {N.}~\bibnamefont {Kim}}, \
  and\ \bibinfo {author} {\bibfnamefont {D.}~\bibnamefont {Waldram}},\ }\href
  {\doibase 10.1103/PhysRevD.63.126001} {\bibfield  {journal} {\bibinfo
  {journal} {Phys. Rev.}\ }\textbf {\bibinfo {volume} {D63}},\ \bibinfo {pages}
  {126001} (\bibinfo {year} {2001})},\ \Eprint
  {http://arxiv.org/abs/hep-th/0012195} {arXiv:hep-th/0012195} \BibitemShut
  {NoStop}%
\bibitem [{\citenamefont {Gaiotto}\ and\ \citenamefont
  {Maldacena}(2012)}]{Gaiotto:2009gz}%
  \BibitemOpen
  \bibfield  {author} {\bibinfo {author} {\bibfnamefont {D.}~\bibnamefont
  {Gaiotto}}\ and\ \bibinfo {author} {\bibfnamefont {J.}~\bibnamefont
  {Maldacena}},\ }\href {\doibase 10.1007/JHEP10(2012)189} {\bibfield
  {journal} {\bibinfo  {journal} {JHEP}\ }\textbf {\bibinfo {volume} {10}},\
  \bibinfo {pages} {189} (\bibinfo {year} {2012})},\ \Eprint
  {http://arxiv.org/abs/0904.4466} {arXiv:0904.4466 [hep-th]} \BibitemShut
  {NoStop}%
\bibitem [{\citenamefont {Anderson}\ \emph {et~al.}(2013)\citenamefont
  {Anderson}, \citenamefont {Beem}, \citenamefont {Bobev},\ and\ \citenamefont
  {Rastelli}}]{Anderson:2011cz}%
  \BibitemOpen
  \bibfield  {author} {\bibinfo {author} {\bibfnamefont {M.~T.}\ \bibnamefont
  {Anderson}}, \bibinfo {author} {\bibfnamefont {C.}~\bibnamefont {Beem}},
  \bibinfo {author} {\bibfnamefont {N.}~\bibnamefont {Bobev}}, \ and\ \bibinfo
  {author} {\bibfnamefont {L.}~\bibnamefont {Rastelli}},\ }\href {\doibase
  10.1007/s00220-013-1675-4} {\bibfield  {journal} {\bibinfo  {journal}
  {Commun. Math. Phys.}\ }\textbf {\bibinfo {volume} {318}},\ \bibinfo {pages}
  {429} (\bibinfo {year} {2013})},\ \Eprint {http://arxiv.org/abs/1109.3724}
  {arXiv:1109.3724 [hep-th]} \BibitemShut {NoStop}%
\bibitem [{\citenamefont {Witten}(1988)}]{Witten:1988ze}%
  \BibitemOpen
  \bibfield  {author} {\bibinfo {author} {\bibfnamefont {E.}~\bibnamefont
  {Witten}},\ }\href {\doibase 10.1007/BF01223371} {\bibfield  {journal}
  {\bibinfo  {journal} {Commun. Math. Phys.}\ }\textbf {\bibinfo {volume}
  {117}},\ \bibinfo {pages} {353} (\bibinfo {year} {1988})}\BibitemShut
  {NoStop}%
\bibitem [{\citenamefont {Bershadsky}\ \emph {et~al.}(1996)\citenamefont
  {Bershadsky}, \citenamefont {Vafa},\ and\ \citenamefont
  {Sadov}}]{Bershadsky:1995qy}%
  \BibitemOpen
  \bibfield  {author} {\bibinfo {author} {\bibfnamefont {M.}~\bibnamefont
  {Bershadsky}}, \bibinfo {author} {\bibfnamefont {C.}~\bibnamefont {Vafa}}, \
  and\ \bibinfo {author} {\bibfnamefont {V.}~\bibnamefont {Sadov}},\ }\href
  {\doibase 10.1016/0550-3213(96)00026-0} {\bibfield  {journal} {\bibinfo
  {journal} {Nucl. Phys. B}\ }\textbf {\bibinfo {volume} {463}},\ \bibinfo
  {pages} {420} (\bibinfo {year} {1996})},\ \Eprint
  {http://arxiv.org/abs/hep-th/9511222} {arXiv:hep-th/9511222} \BibitemShut
  {NoStop}%
\bibitem [{Note1()}]{Note1}%
  \BibitemOpen
  \bibinfo {note} {An orbifold is locally modelled on open subsets of $\protect
  \mathbb {R}^n/\Gamma $, where $\Gamma $ are finite groups.}\BibitemShut
  {Stop}%
\bibitem [{Note2()}]{Note2}%
  \BibitemOpen
  \bibinfo {note} {$\protect \mathbb {WCP}^1_{[n_-,n_+]}$ is a bad orbifold in
  the sense that it is not possible to move to a covering space that is a
  manifold. It does not admit a metric of constant curvature.}\BibitemShut
  {Stop}%
\bibitem [{Note3()}]{Note3}%
  \BibitemOpen
  \bibinfo {note} {The local $D=5$ solutions were independently presented in
  \cite {Kunduri:2006uh}, who also made a local connection with the solutions
  of~\cite {Gauntlett:2006af}.}\BibitemShut {Stop}%
  \bibitem [{\citenamefont {Kunduri}\ \emph {et~al.}(2007)\citenamefont
  {Kunduri}, \citenamefont {Lucietti},\ and\ \citenamefont
  {Reall}}]{Kunduri:2006uh}%
  \BibitemOpen
  \bibfield  {author} {\bibinfo {author} {\bibfnamefont {H.~K.}\ \bibnamefont
  {Kunduri}}, \bibinfo {author} {\bibfnamefont {J.}~\bibnamefont {Lucietti}}, \
  and\ \bibinfo {author} {\bibfnamefont {H.~S.}\ \bibnamefont {Reall}},\ }\href
  {\doibase 10.1088/1126-6708/2007/02/026} {\bibfield  {journal} {\bibinfo
  {journal} {JHEP}\ }\textbf {\bibinfo {volume} {02}},\ \bibinfo {pages} {026}
  (\bibinfo {year} {2007})},\ \Eprint {http://arxiv.org/abs/hep-th/0611351}
  {arXiv:hep-th/0611351} \BibitemShut {NoStop}%
\bibitem [{\citenamefont {Gauntlett}\ \emph {et~al.}(2006)\citenamefont
  {Gauntlett}, \citenamefont {Mac~Conamhna}, \citenamefont {Mateos},\ and\
  \citenamefont {Waldram}}]{Gauntlett:2006af}%
  \BibitemOpen
  \bibfield  {author} {\bibinfo {author} {\bibfnamefont {J.~P.}\ \bibnamefont
  {Gauntlett}}, \bibinfo {author} {\bibfnamefont {O.~A.~P.}\ \bibnamefont
  {Mac~Conamhna}}, \bibinfo {author} {\bibfnamefont {T.}~\bibnamefont
  {Mateos}}, \ and\ \bibinfo {author} {\bibfnamefont {D.}~\bibnamefont
  {Waldram}},\ }\href {\doibase 10.1103/PhysRevLett.97.171601} {\bibfield
  {journal} {\bibinfo  {journal} {Phys. Rev. Lett.}\ }\textbf {\bibinfo
  {volume} {97}},\ \bibinfo {pages} {171601} (\bibinfo {year} {2006})},\
  \Eprint {http://arxiv.org/abs/hep-th/0606221} {arXiv:hep-th/0606221 [hep-th]}
  \BibitemShut {NoStop}%
\bibitem [{\citenamefont {Benini}\ and\ \citenamefont
  {Bobev}(2013)}]{Benini:2012cz}%
  \BibitemOpen
  \bibfield  {author} {\bibinfo {author} {\bibfnamefont {F.}~\bibnamefont
  {Benini}}\ and\ \bibinfo {author} {\bibfnamefont {N.}~\bibnamefont {Bobev}},\
  }\href {\doibase 10.1103/PhysRevLett.110.061601} {\bibfield  {journal}
  {\bibinfo  {journal} {Phys. Rev. Lett.}\ }\textbf {\bibinfo {volume} {110}},\
  \bibinfo {pages} {061601} (\bibinfo {year} {2013})},\ \Eprint
  {http://arxiv.org/abs/1211.4030} {arXiv:1211.4030 [hep-th]} \BibitemShut
  {NoStop}%
\bibitem [{\citenamefont {Gunaydin}\ \emph {et~al.}(1984)\citenamefont
  {Gunaydin}, \citenamefont {Sierra},\ and\ \citenamefont
  {Townsend}}]{Gunaydin:1983bi}%
  \BibitemOpen
  \bibfield  {author} {\bibinfo {author} {\bibfnamefont {M.}~\bibnamefont
  {Gunaydin}}, \bibinfo {author} {\bibfnamefont {G.}~\bibnamefont {Sierra}}, \
  and\ \bibinfo {author} {\bibfnamefont {P.}~\bibnamefont {Townsend}},\ }\href
  {\doibase 10.1016/0550-3213(84)90142-1} {\bibfield  {journal} {\bibinfo
  {journal} {Nucl. Phys. B}\ }\textbf {\bibinfo {volume} {242}},\ \bibinfo
  {pages} {244} (\bibinfo {year} {1984})}\BibitemShut {NoStop}%
\bibitem [{Note4()}]{Note4}%
  \BibitemOpen
  \bibinfo {note} {In general the integral of the curvature for a $U(1)$ gauge
  field on $\protect \mathbb {WCP}^1_{[n_-,n_+]}$ is necessarily $2\pi
  /(n_-n_+)$ times an integer e.g. \cite {Ferrero:2020twa}. This makes the
  gauge field $A$, with flux given by \protect \textup {\hbox {\mathsurround
  \z@ \protect \normalfont (\ignorespaces \ref {Fflux}\unskip \@@italiccorr
  )}}, a spin$^c$ gauge field: $n_--n_+$ is even/odd precisely when $n_-+n_+$
  is even/odd, so that spinor fields with unit charge under $A$ are always
  globally well-defined.}\BibitemShut {Stop}%
\bibitem [{\citenamefont {Ferrero}\ \emph {et~al.}(2020)\citenamefont
  {Ferrero}, \citenamefont {Gauntlett}, \citenamefont {Ipi\~na}, \citenamefont
  {Martelli},\ and\ \citenamefont {Sparks}}]{Ferrero:2020twa}%
  \BibitemOpen
  \bibfield  {author} {\bibinfo {author} {\bibfnamefont {P.}~\bibnamefont
  {Ferrero}}, \bibinfo {author} {\bibfnamefont {J.~P.}\ \bibnamefont
  {Gauntlett}}, \bibinfo {author} {\bibfnamefont {J.~M.~P.}\ \bibnamefont
  {Ipi\~na}}, \bibinfo {author} {\bibfnamefont {D.}~\bibnamefont {Martelli}}, \
  and\ \bibinfo {author} {\bibfnamefont {J.}~\bibnamefont {Sparks}},\
  }\href@noop {} {\  (\bibinfo {year} {2020})},\ \Eprint
  {http://arxiv.org/abs/2012.08530} {arXiv:2012.08530 [hep-th]} \BibitemShut
  {NoStop}%
\bibitem [{\citenamefont {Buchel}\ and\ \citenamefont
  {Liu}(2007)}]{Buchel:2006gb}%
  \BibitemOpen
  \bibfield  {author} {\bibinfo {author} {\bibfnamefont {A.}~\bibnamefont
  {Buchel}}\ and\ \bibinfo {author} {\bibfnamefont {J.~T.}\ \bibnamefont
  {Liu}},\ }\href {\doibase 10.1016/j.nuclphysb.2007.03.001} {\bibfield
  {journal} {\bibinfo  {journal} {Nucl. Phys.}\ }\textbf {\bibinfo {volume}
  {B771}},\ \bibinfo {pages} {93} (\bibinfo {year} {2007})},\ \Eprint
  {http://arxiv.org/abs/hep-th/0608002} {arXiv:hep-th/0608002} \BibitemShut
  {NoStop}%
\bibitem [{Note5()}]{Note5}%
  \BibitemOpen
  \bibinfo {note} {The torus parametrised by $(z,\psi )$ is the same as \cite
  {Gauntlett:2006af}. We have made identifications on $z$ and $\psi $ in the
  opposite order to \cite {Gauntlett:2006af}, so \protect \textup {\hbox
  {\mathsurround \z@ \protect \normalfont (\ignorespaces \ref {delpsi}\unskip
  \@@italiccorr )}} is not the same but $\Delta \psi \Delta z$ is.}\BibitemShut
  {Stop}%
\bibitem [{\citenamefont {Brown}\ and\ \citenamefont
  {Henneaux}(1986)}]{Brown:1986nw}%
  \BibitemOpen
  \bibfield  {author} {\bibinfo {author} {\bibfnamefont {J.}~\bibnamefont
  {Brown}}\ and\ \bibinfo {author} {\bibfnamefont {M.}~\bibnamefont
  {Henneaux}},\ }\href {\doibase 10.1007/BF01211590} {\bibfield  {journal}
  {\bibinfo  {journal} {Commun. Math. Phys.}\ }\textbf {\bibinfo {volume}
  {104}},\ \bibinfo {pages} {207} (\bibinfo {year} {1986})}\BibitemShut
  {NoStop}%
\bibitem [{\citenamefont {Gubser}(1999)}]{Gubser:1998vd}%
  \BibitemOpen
  \bibfield  {author} {\bibinfo {author} {\bibfnamefont {S.~S.}\ \bibnamefont
  {Gubser}},\ }\href {\doibase 10.1103/PhysRevD.59.025006} {\bibfield
  {journal} {\bibinfo  {journal} {Phys. Rev. D}\ }\textbf {\bibinfo {volume}
  {59}},\ \bibinfo {pages} {025006} (\bibinfo {year} {1999})},\ \Eprint
  {http://arxiv.org/abs/hep-th/9807164} {arXiv:hep-th/9807164} \BibitemShut
  {NoStop}%
\bibitem [{\citenamefont {Kim}(2006)}]{Kim:2005ez}%
  \BibitemOpen
  \bibfield  {author} {\bibinfo {author} {\bibfnamefont {N.}~\bibnamefont
  {Kim}},\ }\href {\doibase 10.1088/1126-6708/2006/01/094} {\bibfield
  {journal} {\bibinfo  {journal} {JHEP}\ }\textbf {\bibinfo {volume} {01}},\
  \bibinfo {pages} {094} (\bibinfo {year} {2006})},\ \Eprint
  {http://arxiv.org/abs/hep-th/0511029} {arXiv:hep-th/0511029 [hep-th]}
  \BibitemShut {NoStop}%
\bibitem [{\citenamefont {Gauntlett}\ \emph {et~al.}(2007)\citenamefont
  {Gauntlett}, \citenamefont {Kim},\ and\ \citenamefont
  {Waldram}}]{Gauntlett:2006ns}%
  \BibitemOpen
  \bibfield  {author} {\bibinfo {author} {\bibfnamefont {J.~P.}\ \bibnamefont
  {Gauntlett}}, \bibinfo {author} {\bibfnamefont {N.}~\bibnamefont {Kim}}, \
  and\ \bibinfo {author} {\bibfnamefont {D.}~\bibnamefont {Waldram}},\ }\href
  {\doibase 10.1088/1126-6708/2007/04/005} {\bibfield  {journal} {\bibinfo
  {journal} {JHEP}\ }\textbf {\bibinfo {volume} {04}},\ \bibinfo {pages} {005}
  (\bibinfo {year} {2007})},\ \Eprint {http://arxiv.org/abs/hep-th/0612253}
  {arXiv:hep-th/0612253 [hep-th]} \BibitemShut {NoStop}%
\bibitem [{\citenamefont {Hosseini}\ \emph {et~al.}(2020)\citenamefont
  {Hosseini}, \citenamefont {Hristov}, \citenamefont {Tachikawa},\ and\
  \citenamefont {Zaffaroni}}]{Hosseini:2020vgl}%
  \BibitemOpen
  \bibfield  {author} {\bibinfo {author} {\bibfnamefont {S.~M.}\ \bibnamefont
  {Hosseini}}, \bibinfo {author} {\bibfnamefont {K.}~\bibnamefont {Hristov}},
  \bibinfo {author} {\bibfnamefont {Y.}~\bibnamefont {Tachikawa}}, \ and\
  \bibinfo {author} {\bibfnamefont {A.}~\bibnamefont {Zaffaroni}},\ }\href
  {\doibase 10.1007/JHEP09(2020)167} {\bibfield  {journal} {\bibinfo  {journal}
  {JHEP}\ }\textbf {\bibinfo {volume} {09}},\ \bibinfo {pages} {167} (\bibinfo
  {year} {2020})},\ \Eprint {http://arxiv.org/abs/2006.08629} {arXiv:2006.08629
  [hep-th]} \BibitemShut {NoStop}%
\bibitem [{\citenamefont {Bah}\ \emph {et~al.}(2021)\citenamefont {Bah},
  \citenamefont {Bonetti}, \citenamefont {Minasian},\ and\ \citenamefont
  {Weck}}]{Bah:2020jas}%
  \BibitemOpen
  \bibfield  {author} {\bibinfo {author} {\bibfnamefont {I.}~\bibnamefont
  {Bah}}, \bibinfo {author} {\bibfnamefont {F.}~\bibnamefont {Bonetti}},
  \bibinfo {author} {\bibfnamefont {R.}~\bibnamefont {Minasian}}, \ and\
  \bibinfo {author} {\bibfnamefont {P.}~\bibnamefont {Weck}},\ }\href {\doibase
  10.1007/JHEP02(2021)116} {\bibfield  {journal} {\bibinfo  {journal} {JHEP}\
  }\textbf {\bibinfo {volume} {02}},\ \bibinfo {pages} {116} (\bibinfo {year}
  {2021})},\ \Eprint {http://arxiv.org/abs/2002.10466} {arXiv:2002.10466
  [hep-th]} \BibitemShut {NoStop}%
\end{thebibliography}
%

\end{document}